\documentclass[aps,twocolumn,superscriptaddress]{revtex4-1}
\usepackage{amsbsy,latexsym,amsmath}
\usepackage{amsfonts}
\usepackage{amssymb}
\usepackage[mathscr]{eucal}
\usepackage{epsfig,graphics,graphicx}

\newcommand{\al}{\alpha}

\newcommand{\id}{\openone}
\newcommand{\idd}{{\mathbb I}}

\newcommand{\tr}{{\rm tr}\,}
\newcommand{\trnorm}[1]{\parallel #1 \parallel_1}
\newcommand{\norm}[1]{\parallel #1 \parallel}

\newcommand{\ket}[1]{\left|{#1}\right\rangle}

\newcommand{\braket}[2]{\langle{#1}|{#2}\rangle}
\newcommand{\ketbrad}[1]{\left|{#1}\rangle\!\langle{#1}\right|}
\newcommand{\ketbra}[2]{\left|{#1}\rangle\!\langle{#2}\right|}
\newcommand{\mean}[1]{\langle{#1}\rangle}
\newcommand{\be}{\begin{equation}}
\newcommand{\ee}{\end{equation}}

\newcommand{\half}{\mbox{\footnotesize${1\over2}$}}

\newcommand\xyZ[3]{\mbox{\tiny $\kern-.3em(\!#1\kern-.2em #2\!)\kern-.14em #3\!$}}
\newcommand\Xyz[3]{\mbox{\tiny $\kern-.3em #1\kern-.14em(\! #2\kern-.14em #3\kern-.14em)\!$}}

\renewcommand{\vec}[1]{\mbox{\boldmath$#1$}}
\newcommand{\scrvec}[1]{\mbox{\boldmath\scriptsize$#1$}}

\begin{document}
\title{Robust optimal quantum learning without quantum memory}

\author{G.~Sent\'{i}s}
\author{ J.~Calsamiglia}
\author{R.~Mu\~{n}oz-Tapia}
\affiliation{F\'isica Te\`orica: Informaci\'o i Fen\`omens Qu\`antics, Universitat
Aut\`{o}noma de Barcelona, 08193 Bellaterra (Barcelona), Spain}
\author{E.~Bagan}
\affiliation{F\'isica Te\`orica: Informaci\'o i Fen\`omens Qu\`antics, Universitat
Aut\`{o}noma de Barcelona, 08193 Bellaterra (Barcelona), Spain}
\affiliation{Department of Physics, Hunter College of the City University of New York, 695 Park Avenue, New York, NY 10021, USA}
\affiliation{Physics Department, Brookhaven National Laboratory, Upton, NY 11973, USA}

\begin{abstract}

A quantum learning machine for binary classification of qubit states that does not require quantum memory is introduced and shown to perform with the minimum error rate allowed by quantum mechanics for {\em any} size of the training set. This result is shown to be robust under (an arbitrary amount of) noise and under (statistical) variations in the composition of the training set, provided it is large enough. This machine can be used an arbitrary number of times without retraining. Its required classical memory grows only logarithmically with the number of training qubits, while its excess risk decreases as the inverse of this number, and twice as fast as the excess risk of an ``estimate-and-discriminate" machine, which estimates the states of the training qubits and classifies the data qubit with a discrimination protocol tailored to the obtained estimates.
\end{abstract}

\maketitle

Quantum computers are expected to perform some (classical) computational tasks of practical interest, e.g., large integer factorization, with unprecedented efficiency.
%
Quantum simulators, 
on the other hand, perform tasks of a more ``quantum nature", which cannot be efficiently carried out by a classical computer. Namely, they have the ability to simulate complex  quantum dynamical systems of interest.
%
The need to perform tasks of genuine quantum nature is emerging as individual quantum systems play a more prominent role in labs (and, eventually, in everyday life). Examples include: quantum teleportation, dynamical control of quantum systems, or quantum state identification. Quantum information techniques are already being developed in order
to execute these tasks efficiently. 

This paper is concerned with a simple, yet fundamental instance of quantum state identification. A source produces two unknown pure qubit states with equal probability. A human expert (who knows the source specifications, for instance) classifies a number of $2n$ states produced by this source into two sets of size roughly $n$ (statistical fluctuations of order $\sqrt n$ should be expected) and attaches the labels $0$ and~$1$ to them. We view these~$2n$ states as a training sample, and we set ourselves to find a universal machine that uses this sample to assign the right label to a new unknown state produced by the same source. We refer to this task as quantum classification for short.

Quantum classification can be understood as a {\em supervised quantum learning} problem, as has been noticed by Guta and Kotlowski in their recent work~\cite{guta} (though they use a slightly different setting). Learning theory, more properly named {\em machine} learning theory, is a very active and broad field which roughly speaking deals with algorithms capable of learning from experience~\cite{learning theory}.  Its quantum counterpart~\cite{quantum learning1,quantum learning2,pudenz,bisio,sanders}  not only provides improvements over some classical learning problems but also has a wider range of applicability, which includes the problem at~hand. Quantum learning has also strong links with quantum control theory and is becoming a significant element of the quantum information processing toolbox.


An absolute limit on the minimum error in quantum classification is provided by the so called  optimal programmable discrimination machine~\cite{programmable,Hayashis,He,Sentis}.
In this context, to ensure optimality
one assumes that a fully general
two-outcome joint measurement is performed on {\em both} the~$2n$ training qubits and the qubit we~wish to classify, where the observed outcome determines which of the two labels,~$0$ or~$1$, is assigned to the latter qubit.
Thus, in principle, this assumption implies that in a learning scenario a quantum memory is needed to store the training sample till the very moment we wish to classify the unknown qubit.
%
%
The issue of whether or not the joint measurement assumption can be relaxed has not yet been addressed.
Nor has the issue of how the information left after the joint measurement can be used to classify a second unknown qubit produced by the same source, unless a fresh new  training set (TS) is provided (which may seem unnatural in a learning context).
%

The aim of this paper is to show that for a sizable TS (asymptotically large $n$) the lower bound on the probability of misclassifying the unknown qubit set by programmable discrimination can be attained by first performing a suitable measurement on the TS followed by a Stern-Gerlach type of measurement on the unknown qubit, where forward classical communication is used to control the parameters of the second measurement.
The whole protocol can thus be undersood as a  learning machine (LM), which requires much less demanding assumptions while still having the same accuracy as the optimal programmable discrimination machine. All the relevant information about the TS needed to control the Stern-Gerlach measurement is kept in a {\em classical} memory, thus classification can be executed any time after the learning process is completed. Once trained, this machine can be subsequently used an arbitrary number of times to classify states produced by the same source. Moreover,~this optimal LM is robust under noise, i.e., it still attains optimal performance if the states produced by the source undergo depolarization to any degree.
Interestingly enough, in the ideal scenario where the qubit states are pure and~the TS consists in exactly the same number of copies of each of the two types 0/1  (no statistical fluctuations are allowed) this LM attains the optimal programmable discrimination bound for {\em any} size $2n$ of the TS, not necessarily asymptotically large.


%

At this point it should be noted that LMs without~quantum memory can be naturally assembled from~two quantum information primitives: state estimation and state discrimination. We will refer to these specific constructions as ``estimate-and-discriminate" (E\&D) machines. The protocol they execute is as follows: by performing, e.g., an optimal covariant measurement on the~$n$ qubits in the TS labeled $0$, their state~$|\psi_0\rangle$ is estimated with some accuracy, and likewise the state~$|\psi_1\rangle$ of the other~$n$ qubits that carry the label~$1$ is characterized. 
This classical information is stored and subsequently used to discriminate an unknown qubit state. 
It will be shown that the excess risk (i.e., excess average error over classification when the states $|\psi_0\rangle$ and $|\psi_1\rangle$ are perfectly known) of this protocol is twice  that of the optimal LM.
The fact that the E\&D machine is suboptimal means that the kind of information retrieved from the TS and stored in the classical memory of the optimal LM is specific to the classification problem at hand, and that the machine itself is more than the mere assemblage of well known protocols.

We will first present our results for the ideal scenario where states are pure and no statistical fluctuation in the number of copies of each type of state is allowed. The effect of these fluctuations and the robustness of the LM optimality against noise will be postponed to the end of the section. 

\section*{Results}

\textbf{Programmable machines}. Before presenting our results, let us summarize what is known about optimal machines for programmable discrimination. This will also allow us to introduce our notation and conventions. Neglecting statistical fluctuations, the TS of size $2n$ is given by a state pattern of the form $[\psi_0^{\otimes n}]\otimes[\psi_1^{\otimes n}]$, where the shorthand notation $[\,\cdot\,]\equiv|\,\cdot\,\rangle\langle\,\cdot\,|$ will be used throughout the paper, and where no knowledge about the actual states $|\psi_0\rangle$ and $|\psi_1\rangle$ is assumed (the figure of merit will be an average over {\em all} states of this form). The qubit state that we wish to label (the \emph{data} qubit) belongs either to the first group (it is $[\psi_0]$)  or to the second one (it is $[\psi_1]$). Thus, the optimal machine must discriminate between the two possible states: either $\varrho^n_0=[\psi_0^{\otimes (n+1)}]_{AB}\otimes[\psi_1^{\otimes n}]_{C}$, in which case it should output the label $0$, or $\varrho^n_1=[\psi_0^{\otimes n}]_{A}\otimes[\psi_1^{\otimes (n+1)}]_{BC}$, in which case the machine should output the label $1$. Here and when needed for clarity, we name the three subsystems involved in this problem~$A$,~$B$ and~$C$, where $AC$ is the TS and~$B$ is the data qubit. In order to discriminate~$\varrho^n_0$ from~$\varrho^n_1$, a joined two-outcome measurement, independent of the actual states $|\psi_0\rangle$ and $|\psi_1\rangle$, is performed on all~$2n+1$ qubits. Mathematically, it is  represented by a positive operator valued  measure (POVM) ${\mathscr E}=\{E_0,E_1=\openone-E_0\}$. The minimum average error probability of the quantum classification process is given by $P_{\rm e}=(1-\Delta/2)/2$, where $\Delta=2\max_{E_0}\int d\psi_0\,d\psi_1\; \tr\left[\left(\varrho^n_0-\varrho^n_1\right)E_0\right]$.
This average can be cast as a $\rm SU(2)$ group integral and, in turn, readily computed using Schur's lemma to give
\begin{equation}
\Delta=2\max_{E_0}\tr\left[\left(\sigma^n_0-\sigma^n_1\right)E_0\right]=\parallel\sigma^n_0-\sigma^n_1\parallel_1
\label{Delta},
\end{equation}
where ${\parallel\cdot\parallel_1}$ is the trace norm and $\sigma^n_{0/1}$ are average states defined as
\begin{eqnarray}\label{sigma states}
\sigma^n_0&=&\frac{\openone_{n+1}\otimes\openone_n}{d_{n+1}d_{n}}=\frac{\openone_{AB}\otimes\openone_C}{d_{AB}d_{C}}  , \nonumber \\
\sigma^n_1&=&\frac{\openone_{n}\otimes\openone_{n+1} }{d_{n}d_{n+1}} =\frac{\openone_{A}\otimes\openone_{BC}}{d_{A}d_{BC}}.
\end{eqnarray}
%
%
In this paper~$\openone_m$ stands for the projector on the fully symmetric invariant subspace of $m$ qubits, which has dimension $d_{m}=m+1$. Sometimes, it turns out to be more convenient to use the subsystem labels, as on the right of~(\ref{sigma states}). The maximum in~(\ref{Delta}) is attained by choosing $E_0$ to be the projector onto the positive part of $\sigma^n_0-\sigma^n_1$. 

The right-hand side of~(\ref{Delta}) can be computed by switching to the total angular momentum basis, $\{|J,M\rangle\}$, where $\half\le J\le n+\half$ and $-J\le M\le J$ (an additional label may be required to specify the way subsystems couple to give $J$; see below). In this (Jordan) basis~\cite{He} the problem simplifies significantly, as it reduces to pure state discrimination~\cite{pure state discr} on each subspace corresponding to a possible value of the total angular momentum $J$ and magnetic number~$M$.  
%
%
By~writing the various values~of the total angular momentum as~$J=k+\half$, the final answer takes the form~{\cite{Sentis}}:
\begin{equation}
P^{\rm opt}_{\rm e}={1\over2}-\frac{1}{d_n^2 d_{n+1}} \sum_{k=0}^n k \sqrt{d_n^2-k^2} .
 \label{OptDisc}
\end{equation}
A simple asymptotic expression for large $n$ can be computed using Euler-Maclaurin's summation formula. After some algebra one obtains
\begin{equation}\label{progr asymp}
P^{\rm opt}_e\simeq {1\over 6}+{1\over3n} .
\end{equation}
The leading order ($1/6$) coincides with the average error probability $\int d\psi_0\,d\psi_1 \,p^{\rm opt}_e(\psi_0,\psi_1)$, where $p^{\rm opt}_e(\psi_0,\psi_1)$ is the minimum error in discrimination between the two {\em known} states $\ket{\psi_0}$ and $\ket{\psi_1}$.
\\

\textbf{Learning machines}. The formulas above give an absolute lower bound to the error probability  that can be physically attainable. We wish to show that this bound can actually be attained by a learning machine that uses a classical register to store all the relevant information obtained in the  learning process regardless the size, $2n$, of the TS.
A first hint that this may be possible is that the optimal measurement~${\mathscr E}$ can be shown to have positive partial transposition with respect to the partition TS/data qubit. Indeed this is a necessary condition for any measurement that consists of a local POVM on the TS whose outcome is fed-forward to a second POVM on the data qubit. This class of one-way adaptive measurement can be characterized as:
\begin{equation}
E_0=\sum_\mu L_\mu\otimes D_\mu,\quad
E_1=\sum_\mu L_\mu\otimes (\openone_1-D_\mu),
\label{LM POVM}
\end{equation}
where the positive operators  $L_\mu$ ($D_\mu$) act on the Hilbert space of the TS (data qubit we wish to classify), and $\sum_\mu L_\mu=\openone_n\otimes\openone_n$.  The POVM ${\mathscr L}=\{L_\mu\}$ represents the learning process, and the parameter $\mu$, which a priori may be discrete or continuous, encodes the information gathered in the measurement and required at the classification stage. For each possible value of~$\mu$, ${\mathscr D}_\mu=\{D_\mu,\openone_1-D_\mu\}$ defines the measurement on the data qubit, whose two outcomes represent the classification decision.  Clearly, the size of the required classical memory will be determined by the information content of the random variable~$\mu$.
\\

\textbf{\boldmath Covariance and structure of $\mathscr{L}$}. 
We will next prove that the POVM $\mathscr L$, which extracts the relevant information from the TS, can be chosen to be covariant. This will also shed some light on the physical interpretation of the classical variable $\mu$.
The states (\ref{sigma states}) are by definition invariant under a rigid rotation acting on subsystems $AC$ and $B$, of the form $U=U_{AC} \otimes u$, where throughout this paper, $U$ stands for an element of the appropriate representation of SU(2), which should be obvious by context (in this case $U_{AC}=u^{\otimes 2n}$, where $u$ is in the fundamental representation). Since
$
\tr ( E_0 \sigma^n_{0/1} ) = \tr (E_0 U^{\dagger} \sigma^n_{0/1} U ) = \tr (U E_0 U^{\dagger} \sigma^n_{0/1} )
$, the positive operator $UE_0U^\dagger$ gives the same error probability as $E_0$ for {\em any} choice of $U$ [as can be seen from, e.g., Eq.~(\ref{Delta})]. The same property thus holds for their average over the whole SU(2) group 
$
\bar E_0 = \int du \, U E_0 U^{\dagger}
$, 
which is invariant under rotations, and where $du$ denotes the SU(2) Haar measure. By further exploiting rotation invariance (see Sec.~Methods for full details) $\bar E_0$ can be written as
 %
\begin{equation}\label{LM covariant POVM}
\bar E_0=\int du \, \left(U_{AC}\,{\Omega}\,U^\dagger_{AC}\right)\otimes \left(u [\,\uparrow\,] u^{\dagger}\right)
\end{equation}
%
%
for some positive operator $\Omega$, where we use the short hand notation $[\,\uparrow\,]\equiv\ketbrad{\half,\half}$. Similarly, the second POVM element can be chosen to be an average,~$\bar E_1$, of the form~(\ref{LM covariant POVM}), with $[\,\downarrow\,]\equiv\ketbrad{\half,-\half} $ instead of~\mbox{$[\,\uparrow\,]$}. We immediately recognize $\bar{\mathscr E}=\{\bar E_0,\bar E_1\}$ to be of the form~(\ref{LM POVM}), where $u$,  $L_u\equiv U_{AC}\,{\Omega}\,{U}_{AC}^{\dagger}$ and $D_u\equiv u[\,\uparrow\,]u^\dagger$ play the role of~$\mu$, $L_\mu$ and~$D_\mu$ respectively. 
Hence,  w.l.o.g. we can choose ${\mathscr L}=\{U_{AC}\,{\Omega}\,{U}_{AC}^{\dagger}\}_{\rm SU(2)}$, which is a covariant POVM with seed~$\Omega$.
Note that $u$ entirely defines the Stern-Gerlach measurement, ${\mathscr D}_u=\{u[\,\uparrow\,]u^\dagger, u[\,\downarrow\,]u^\dagger \}$, 
i.e., $u$ specifies the direction along which the Stern-Gerlach has to be oriented. This is the relevant information that has to be retrieved from the TS and kept in the classical memory of the LM.
%
%

Covariance has also implications on the structure of~${\Omega}$. 
In Sec.~Methods, we show that this seed can always be written as
\begin{equation}\label{omega}
{\Omega} = \sum_{m=-n}^n {\Omega}_m \, ; \quad {\Omega}_m \geqslant 0 \, ,
\end{equation}
where
\begin{equation}\label{omega_m}
\sum_{m=-j}^j \langle j,m|\Omega_m| j,m\rangle= 2j+1 ,\quad 0\le j\le n,
\end{equation}
%
%
and $j$ ($m$) stands for the total angular momentum~$j_{AC}$ (magnetic number $m_{AC}$) of the qubits in the~TS. In other words,~the seed is a direct sum of operators with a well defined magnetic number. As a result, we can interpret that ${\Omega}$ points along the $z$-axis. The constrain~(\ref{omega_m}) ensures~that $\mathscr L$ is a resolution of the identity.

To gain more insight into the structure of~$\Omega$, we trace subsystems~$B$ in the definition of $\Delta$, given by the first equality in Eq.~(\ref{Delta}). For the covariant POVM~(\ref{LM covariant POVM}),  rotational invariance enables us to express this quantity as 
%
\begin{equation}
\Delta^{\mathrm{\!LM}}\!=\! 2\max_\Omega \tr\!\!\left\{(\sigma^n_0\!-\!\sigma^n_1) \Omega\!\otimes\![\,\uparrow\,]\right\}\!=\! 2\max_\Omega \tr\! (\Gamma_\uparrow \Omega),
\label{Delta SDP}
\end{equation}
where we have defined 
\begin{equation}\label{def GammaUp}
\Gamma_\uparrow\!=\!\tr_{\!B}\{[\,\uparrow\,] (\sigma^n_0-\sigma^n_1)\}
\end{equation}
(the two resulting terms in the right-hand side are the post-measurement states of~$AC$ conditioned to the outcome~$\uparrow$ after the Stern-Gerlach measurement ${\mathscr D}_{\scrvec z}$ is performed on~$B$)
 and the maximization is over valid seeds (i.e., over positive operators~$\Omega$ such that $\int du\, U_{AC}\,\Omega\, U^\dagger\kern-.3em{}_{AC}=\id_{AC}$). We calculate~$\Gamma_\uparrow$ in Sec.~Methods. The resulting expression can be cast in the simple and transparent form 
%
%
\begin{equation}
\Gamma_\uparrow={\hat J^{A}_z-\hat J^{C}_z\over d_n^2 d_{n+1}},
\label{JA-JC}
\end{equation}
%
where $\hat J^{A/C}_z$ is the $z$ component of the total angular momentum operator acting on subsystem $A/C$,
i.e., on the training qubits to which the human expert assigned the label~0/1.
Eq.~(\ref{JA-JC}) suggests that the optimal $\Omega$ should project on the subspace of $A$ ($C$) with maximum (minimum) magnetic number, which implies that $m_{AC}=0$. An obvious candidate is 
\begin{equation}\label{seed}
\Omega=[\phi^0], \quad
\ket{\phi^0} = \sum_{j=0}^n \sqrt{2j+1} \ket{j,0} .
\end{equation}
Below we prove that indeed this seed generates the optimal~LM~POVM. 
\\

\textbf{Optimality of the LM}. 
We now prove our main result: the POVM $\bar{\mathscr E}=\{\bar{E}_0,\bar{E}_1\}$, generated from the seed state in Eq.~(\ref{seed}), gives an error probability $P_{\rm e}^{\rm LM}=(1-\Delta^{\rm LM}/2)/2$  equal to the minimum error probability~$P^{\rm opt}_{\rm e}$ of the optimal programmable discriminator,~Eq.~(\ref{OptDisc}). It is, therefore, optimal and, moreover, it attains the absolute minimum allowed by quantum physics. 

The proof goes as follows. From the very definition of error probability,
\begin{equation}
P_{\rm e}^{\rm LM} = \frac{1}{2} \left(\tr \sigma^n_1\bar E_0+\tr \sigma^n_0\bar E_1\right),
\end{equation}
 we have
 \begin{equation}
P_{\rm e}^{\rm LM}\!\!=\!\!
 {\tr\!\!\left(\!\openone_{\!A}\!\!\otimes\!\!\openone_{\!BC} [\phi^0]\!\otimes\!{[\uparrow]}\right)\!+\!
 \tr\!\!\left(\!\openone_{\!AB}\!\otimes\!\!\openone_{\!C} [\phi^0]\!\otimes\![\downarrow]\right)\over 2 d_n d_{n+1}} ,
\end{equation}
where we have used rotational invariance. We can further simplify this expression by writing it as 
\begin{equation}
P_{\rm e}^{\rm LM}\! \!=\!{
\norm{\!\!\id_A\!\! \otimes\!\! \id_{BC} \!\ket{\phi_0}\ket{\uparrow}\!\!}^{\,2}
\!\! +\! \norm{\!\!\id_{AB}\! \otimes\!\! \id_C \ket{\phi_0}\!\ket{\downarrow}\!\!}^{\,2} 
\over 2 d_n d_{n+1}} .
\label{P^LM norm}
\end{equation}
%
%
%
%
To compute the projections inside the norm signs
we first write $|\phi^0\rangle|\!\!\uparrow\,\rangle$ ($|\phi^0\rangle|\!\!\downarrow\,\rangle$ will be considered below) in the total angular momentum basis $|J,M\rangle_{\xyZ ACB}\,$, where the attached subscripts remind us how subsystems $A$, $B$ and~$C$ are  both \emph{ordered} and {\em coupled} to give the total angular momentum~$J$ (note that a permutation of subsystems, prior to fixing the coupling, can only give rise to a global phase, thus not affecting the value of the norm we wish to compute). This is a trivial task since $|\phi^0\rangle|\!\!\uparrow\rangle\equiv|\phi^0\rangle_{AC}|\!\!\uparrow\rangle_B$, i.~e., subsystems are ordered and coupled as the subscript $(AC)B$ specifies, so we just need the Clebsch-Gordan coefficients 
\begin{equation}
\braket{j\pm\half,\half}{j,0;\half,\half}=\pm\sqrt{\frac{j+\frac{1}{2}\pm\frac{1}{2}}{2j+1}} .
\end{equation}

The projector $\openone_{A}\otimes\openone_{BC}$, however, is naturally written as
$
\openone_{A}\otimes\openone_{BC}=\sum_{J,M}|J,M\rangle_{\Xyz ACB}\langle J,M|
$.
This basis differs from that above in the coupling of the subsystems. To compute the projection $\openone_A\!\otimes\!\openone_{BC} |\phi^0\rangle|{\uparrow\rangle}$ we only need to know the overlaps between the two bases ${{}_{\Xyz ACB}\kern-.1em\langle J,M|J,M\rangle_{\xyZ ACB}}$. Wigner's 6j-symbols provide this information as a function of the angular momenta of the various subsystems (the overlaps are computed explicitly in Sec.~Methods).
 
Using the Clebsch-Gordan coefficients and the overlaps between the two bases, it is not difficult  to obtain
\begin{equation}
 \id_{\!A}\!\otimes\!\id_{\!BC} |\phi^0\rangle|\!\!\uparrow\rangle\!\!=\!\!
 \sum_{j=1}^{n+1}\!
\!\sqrt{\! j} {\sqrt{d_n\!+\!j}\!-\!\sqrt{d_n\!-\!j}\over\sqrt2 d_n}|j\!-\!\half\!,\!\half\rangle_{\Xyz ACB},
\label{projection}
 \end{equation}
 %
An identical expression can be obtained for $\openone_{AB}\otimes\openone_C \ket{\phi^0}\ket{\downarrow}$ in the basis $|J,M\rangle_{\xyZ BAC}\,$. To finish the proof, we compute the norm squared of~(\ref{projection}) and 
substitute in~(\ref{P^LM norm}).
 It is easy to check that this gives the expression of the error probability 
in~(\ref{OptDisc}), i.e.,~$P_e^{\rm LM}=P^{\rm opt}_e$. 
 \\

\textbf{Memory of the LM}. Let us go back to the POVM condition, specifically to the minimum number of unitary transformations needed to ensure that, given a suitable discretization $\int du \to \sum_\mu p_\mu$ of (\ref{LM covariant POVM}), $\{p_\mu U_\mu[\,\phi^0\,]U^\dagger_\mu\}$ is a resolution of the identity for arbitrary~$n$. 
This issue is addressed in~\cite{BBM-T}, where an explicit algorithm for constructing finite POVMs, including the ones we need here, is given.  From the results there, we can bound the minimum number of outcomes of $\mathscr L$ by $2(n+1)(2n+1)$.
This figure is important because its binary logarithm gives an upper bound to the minimum memory required. 
We see that it  grows at most logarithmically with the size of the TS.\\

\textbf{E\&D machines}. E\&D machines can be discussed within this very framework, as they are particular instances of LMs. In this case the POVM~$\mathscr L$ has the form
$L_{\al i} =M_\al\otimes M'_i$, where ${\mathscr M}=\{M_\al\}$ and ${\mathscr M}'=\{M'_i\}$ are themselves POVMs on the TS subsystems $A$ and~$C$ respectively. The role of~${\mathscr M}$ and~${\mathscr M}'$ is to estimate (optimally) the qubit states in these subsystems~\cite{Holevo}. The measurement on~$B$ (the data qubit) now depends on the pair of outcomes of~${\mathscr M}$ and~${\mathscr M}'$: ${\mathscr D}_{\al i}=\{D_{\al i},\openone_1-D_{\al i}\}$. It performs standard one-qubit discrimination according to the two pure-state specifications, say, the unit Bloch vectors $\vec s_0^\al$ and $\vec s_1^i$, estimated with ${\mathscr M}$ and~${\mathscr M}'$. 
In this section,
we wish to show that E\&D machines perform worse than the optimal LM. 

We start by tracing subsystems~$AC$ in Eq.~(\ref{Delta}), which for E\&D reads
\begin{equation}
\Delta^{\rm E\&D}=2\max_{{\mathscr M}, {\mathscr M}'}\tr_{\!\! B} \max_{\{{\mathscr D}_{\alpha i}\}}\tr_{\!\!AC}[(\sigma^n_0-\sigma^n_1)E_0].
\end{equation}
 If we write $\Delta^{\rm E\&D}=\max_{\mathscr {\mathscr M},{\mathscr M}'}\Delta_{{\mathscr M},{\mathscr M}'}$, we have
\begin{equation}
\Delta_{{\mathscr M},{\mathscr M}'}=\sum_{\al i}p_\al p'_i |\vec r_0^\al-\vec r_1^i| ,
\label{E+D}
\end{equation} 
where $\vec r_0^\al$ and $\vec r_1^i$ are the Bloch vectors of the data qubit states 
\begin{equation}\label{conditioned}
\rho^\alpha_0\!=\!{1\over p_\alpha}\tr_{\!\!A}\!\!\left(\!{\id^{AB}_{n+1}\over d_{n+1}}M_\alpha\!\!\right),
\quad
\rho^i_1\!=\!{1\over p'_i}\tr_{\!\!C}\!\!\left(\!{\id^{BC}_{n+1}\over d_{n+1}}M'_i\!\!\right),
\end{equation}
conditioned to the outcomes~$\alpha$ and~$i$ respectively,  
and~$p_\alpha=d^{-1}_n\tr M_\alpha$, $p'_i=d^{-1}_n\tr M'_i$ are their probabilities. 
We now recall that optimal estimation necessarily requires that all elements of $\mathscr M$ must be of the form $M_\alpha=c_\alpha U_\alpha[\psi^0] U^\dagger_\alpha$, where~$|\psi^0\rangle=|\mbox{\footnotesize${n\over2}$},\mbox{\footnotesize${n\over2}$}\rangle$, $c_\alpha>0$, and~$\{U_\alpha\}$ are appropriate SU(2) rotations  (analogous necessary conditions are required for~${\mathscr M}'$)~\cite{DBE}. Substituting in Eq.~(\ref{conditioned}) we obtain 
$
p_\alpha={c_\alpha/d_n}$,  %
and
\begin{equation}
u^\dagger_\alpha\rho^\alpha_{0} u_\alpha={1\over d_{n+1}}\left(d_n[\,\uparrow\,]+[\,\downarrow\,]\right) 
\end{equation}
(a similar expression holds for $\rho^i_1$).
This means that the Bloch vector of the data qubit conditioned to outcome~$\alpha$ is proportional to $\vec s^\alpha_0$ (the Bloch vector of the corresponding estimate) and is shrunk by a factor $n/d_{n+1}=n/(n+2)=\eta$. Note in passing that the shrinking factor~$\eta$ is independent of the measurements, provided it is optimal. 

Surprisingly at first sight, POVMs that are optimal, and thus equivalent, for estimation may lead to different minimum error probabilities. In particular, the continuous covariant POVM is outperformed  in the problem at hand by those with a finite number of outcomes. 
Optimal POVMs with few outcomes enforce large angles between the estimates~$\vec s_0^\al$ and $\vec s_1^i$, and thus between $\vec r_0^\al$ and $\vec r_1^i$ ($\pi/2$ in the $n=1$ example below). This translates into increased discrimination efficiency, as shown by~(\ref{E+D}), without compromising the quality of the estimation itself. Hence the orientation of  ${\mathscr M}$ relative to ${\mathscr M}'$ (which for two continuous POVMs does not even make sense) plays an important role, as it does the actual number of outcomes. 
With an increasing size of the TS, the optimal estimation POVMs require also a larger number of outcomes and the angle between the estimates decreases in average, since they tend to fill the 2-sphere isotropically. Hence, the minimum error probability is expected to approach that of two continuous POVMs. This is supported by numerical calculations. 
The problem of finding the optimal E\&D machine for arbitrary $n$ appears to be a hard one and is currently under investigation. Here we will give the absolute optimal E\&D machine for $n=1$ and, also, we will compute the minimum error probability for both~$\mathscr M$ and~${\mathscr M}'$ being the continuous POVM that is optimal for estimation. The later, as mentioned,  is expected to attain the optimal E\&D error probability asymptotically.

We can obtain an upper bound on~(\ref{E+D}) by applying the Schwarz inequality.
We~readily find that 
%
\begin{eqnarray}
\Delta_{{\mathscr M},{\mathscr M}'} &\leqslant& \sqrt{\sum_{\al i} p_\al p'_i|\vec r_0^\alpha-\vec r_1^i|^2} \nonumber \\
&=&\sqrt{\sum_{\al} p_\al|\vec r_0^\alpha|^2+\sum_{i}p'_i |\vec r_1^i|^2} ,
\end{eqnarray}
where we have used that $\sum_\al p_\alpha \vec r_0^\alpha=\sum_i p'_i \vec r_1^i=0$, as follows from the POVM condition on $\mathscr M$ and ${\mathscr M}'$. The maximum norm of $\vec r^\alpha_0$ and $\vec r^i_1$ 
is bounded by $1/3$ [the shrinking factor $\eta$ for $n=1$]. 
Thus 
\begin{equation}
\Delta_{{\mathscr M},{\mathscr M}'} \leqslant \sqrt{2}/3<1/\sqrt3=\Delta^{\rm LM},
\label{Delta n=1}
\end{equation}
where
the value of $\Delta^{\rm LM}$ can be read off from Eq.~(\ref{OptDisc}).
The E\&D bound $\sqrt2/3$ is attained by the choices \mbox{$M_{\uparrow/\downarrow}=[\,\uparrow\!\!/\!\!\downarrow\,]$} and $M'_{+/-}=[+\!/\!-]$, where we have used the definition~$|\pm\rangle=(|\!\uparrow\,\rangle\pm|\!\downarrow\,\rangle)/\sqrt2$.

For arbitrary $n$, a simple expression for the error probability can be derived in the continuous POVM case, ${\mathscr M}={\mathscr M}'\!=\{{d_n} U_{\scrvec s}[\,\psi^0\,]U^\dagger_{\scrvec s}\}_{\scrvec s\in{\mathbb S}^2}$, where 
$\vec s$ is a unit vector (a point on the 2-sphere ${\mathbb  S}^2$) and~$U_{\scrvec s}$ is the representation of the rotation that takes the unit vector along the $z$-axis,~$\vec z$, into~$\vec s$. Here $\vec s$ labels the outcomes of the measurement and thus plays the role of~$\alpha$ and $i$. 
The continuous version of~(\ref{E+D}) can be easily computed to be
\begin{equation}
\Delta^{\rm E\&D}=\eta  \int d\vec s\,|{\vec z}-\vec s|=\frac{4n}{3(n+2)}\,.
\end{equation}
Asymptotically, we have $P_{\rm e}^{\rm E\&D}=1/6+2/(3n)+\dots$. Therefore, the excess risk, which we recall is the difference between the average error probability of the machine under consideration and that of the optimal discrimination protocol for {\em known} qubit states 
 ($1/6$), is $R^{\rm E\&D}=2/(3n)+\dots$. This is twice the excess risk of the optimal programmable machine and the optimal LM, which can be read off from Eq.~(\ref{progr asymp}): 
 \begin{equation}\label{ex risk}
 R^{\rm LM}=R^{\rm opt}={1\over 3n}+\dots.
 \end{equation}
  For $n=1$, Eq.~(\ref{Delta n=1}) leads to $R^{\rm E\&D}=(4-\sqrt2)/12$.
 This value is already~$15\%$ larger than excess risk of the optimal LM:~$R^{\rm LM}=(4-\sqrt3)/12$. \\



\textbf{Robustness of LMs}.
So far we have adhered to the simplifying assumptions that the two types of states produced by the source are pure and, moreover, exactly equal in number. Neither of these two assumptions is likely to hold in practice, as both, interaction with the environment, i.e., decoherence and noise, and statistical fluctuations in the numbers of states of each type, will certainly take place. Here we prove that the performance of the optimal LM  is not altered by these effects in the asymptotic limit of large TS. More precisely, the excess risk of the optimal~LM remains equal to that of the optimal programmable discriminator to leading order in $1/n$ when noise and statistical fluctuations are taken into account.

Let us first consider the impact of noise,  which we will assume isotropic and uncorrelated. Hence, instead of producing $[\psi_{0/1}]$, the source produces copies of
%
%
%
\begin{equation}
\rho_{0/1} = r [\psi_{0/1}] + (1-r) \frac{\id}{2},\quad 0<r\le1.
\label{rho mixed}
\end{equation}
%
In contrast to the pure qubits case, where $[\psi^{\otimes n}_{0/1}]$ belongs to the fully symmetric invariant subspace of maximum angular momentum $j=n/2$, the state of $A/C$ is now a full-rank matrix of the form $\rho^{\otimes n}_{0/1}$. Hence, it has projections on all the orthogonal subspaces
$\mathscr{S}_{j}\otimes{\mathbb C}^{\nu^n_j}$, where~$\mathscr{S}_{j}={\rm span}(\{\ket{j,m}\}_{m=-j}^j)$ and ${\mathbb C}^{\nu^n_j}$ is the multiplicity space of the representation with total angular momentum $j$ (see Sec.~Methods for a formula of the multiplicity~$\nu^n_j$), and~$j$ is in the range from~$0$~($1/2$) to $n/2$ if $n$ is even (odd). Therefore $\rho^{\otimes n}_{0/1}$ is block-diagonal in the total angular momentum eigenbasis. The multiplicity space~${\mathbb C}^{\nu^n_j}$ carries the label of the $\nu^n_j$ different equivalent representations of given~$j$, which arise from the various ways the individual qubits can couple to produce total angular momentum~$j$. 
For permutation invariant states (such as~$\rho^{\otimes n}_{0/1}$), this has no physical relevance and the only effect of ${\mathbb C}^{\nu^n_j}$ in calculations is through its dimension $\nu^n_j$. Hence,  the multiplicity space will be dropped throughout this paper. 

The average states now become a direct sum of the form
\begin{eqnarray}
 \int d\psi_0\,d\psi_1\, \rho_0^{\otimes (n+1)} \otimes \rho_1^{\otimes n} &=& \sum_\xi p^n_\xi \sigma^n_{0,\xi} \label{sigmaxi1} , \\
 \int d\psi_0\,d\psi_1\, \rho_0^{\otimes n} \otimes \rho_1^{\otimes (n+1)} &=&\sum_{\xi} p^n_{\xi}\sigma^n_{1,\xi}   \label{sigmaxi2} ,
\end{eqnarray}
where we use the shorthand notation $\xi=\{ j_A, j_C\}$ 
[each angular momentum ranges from $0$ ($1/2$) to $n/2$ for $n$ even (odd)], and
$p^n_\xi=p^n_{j_A}p^n_{j_C}$ is the probability of any of the two average states projecting on the block labeled~$\xi$. 
%
%
%
%
%
Hence, 
\begin{equation}\label{delta LM sum xi}
\Delta^{\rm LM}= \sum_\xi p^n_\xi \parallel \sigma^n_{0,\xi} -\sigma^n_{1,\xi} \parallel_1 .
\end{equation}
%
The number of terms in Eq.~(\ref{delta LM sum xi}), is $[(2n+3\pm 1)/4]^2$ for even/odd $n$.
It grows quadratically with $n$, in contrast to the pure state  case for which there is a single contribution corresponding to $j_A=j_C=n/2$. In the asymptotic limit of large $n$, however, a big simplification arises because of the following results: for each $\xi$ of the form $\xi=\{j,j\}$ \mbox{($j_A=j_C=j$)}, the following relation holds (see Sec.~Methods)
\begin{equation}\label{ja=jc}
{\sigma^n_{0,\xi} -\sigma^n_{1,\xi} } ={r\langle\hat{J}_z\rangle_j\over j} \left({\sigma^{2j}_{0}-\sigma^{2j}_{1}}\right) ,
\end{equation}
where $\sigma^{2j}_{0/1}$ are the average states~(\ref{sigma states}) for a number of~$2j$ \emph{pure} qubits. Here~$\langle\hat{J}_z\rangle_j$ is the expectation value restricted to~$\mathscr{S}_{j}$ of the \mbox{$z$-component} of the angular momentum in the state $\rho^{\otimes n}$, where $\rho$ has Bloch vector $r\vec{z}$. Eq.~(\ref{ja=jc}) is an exact algebraic identity that holds for any value of $j$, $n$ and $r$ (it bears no relation whatsoever to measurements of any kind).  The second result is that for large~$n$, both~$p^n_{j_A}$ and~$p^n_{j_C}$ become continuous probability distributions, $p_n(x_A)$ and $p_n(x_C)$, where $x_{A/C}=2 j_{A/C}/n\in[0,1]$. Asymptotically, they approach Dirac delta functions peaked at~\mbox{$x_A=x_C=r$} (see Sec.~Methods). Hence, the only relevant  contribution to~$\Delta^{\rm LM}$ comes from~$\xi=\{rn/2,rn/2\}$. It then follows that in the asymptotic limit
\begin{equation}\label{s-s asymp}
\sum_\xi p^n_\xi\left(\sigma^n_{0,\xi}-\sigma^n_{1,\xi}\right)\simeq {2\langle\hat{J}_z\rangle_{rn/2}\over n}
\left(\sigma^{rn}_{0}-\sigma^{rn}_{1}\right).
\end{equation}
%
%
Hence, mixed-state quantum classification using a TS of~size $2n$ is equivalent to its {\em pure}-state  version for a TS of size $2nr$, provided $n$ is asymptotically large. In particular, our proof of optimality above also holds for {\em arbitrary}~$r\in(0,1]$ if the TS is sizable enough, and~\mbox{$R^{\rm LM}\simeq R^{\rm opt}$}.
%
This result is much stronger than robustness against decoherence, which only would require optimality for values of $r$ close to unity. 

{}From Eqs.~(\ref{delta LM sum xi}) and~(\ref{s-s asymp}) one can easily compute~$\Delta^{\rm LM}$ for arbitrary $r$ using that~\cite{BEJRE} $\langle\hat{J}_z\rangle_j\simeq j-(1-r)/(2r)$ up to exponentially vanishing terms. The trace norm of~$\sigma^{rn}_{0}-\sigma^{rn}_{1}$ can be retrieved from, e.g., Eq.~(\ref{ex risk}). For~$rn$ {\em pure} qubits one has $\trnorm{\sigma^{rn}_{0}-\sigma^{rn}_{1}}\simeq (4/3)[1-1/(rn)]$. After some trivial algebra we obtain 
\begin{equation}
P^{\rm LM}_e={1\over2}-{r\over3}+{1\over3rn}+o(n^{-1})
\end{equation}
for the error probability, in agreement with the optimal programmable machine value given in~\cite{Sentis}, as claimed above. This corresponds to an 
excess risk of
\begin{equation}\label{rlm=ropt}
R^{\rm LM}={1\over3rn}+o(n^{-1}) = R^{\rm opt}.
\end{equation}

In the non-asymptotic case, the sum in Eq.~(\ref{delta LM sum xi}) is not restricted to $\xi=\{j,j\}$ and the calculation of the excess risk becomes very involved. Rather than attempting to obtain an analytical result, for small training samples we have resorted to a numerical optimization.
%
We first note that
Eqs.~(\ref{omega}) through (\ref{JA-JC}) define a \emph{semidefinite programming} optimization problem (SDP), for which very efficient numerical algorithms have been developed~\cite{SDP}.  In this framework, one maximizes the objective function~$\Delta^{\rm LM}$ [second equality in Eq.~(\ref{Delta SDP})] of the SDP variables~${\Omega}_m\ge0$, subject to the linear condition (\ref{omega_m}).
We use this approach to  compute the error probability, or equivalently, the excess risk of a LM for mixed-state quantum classification of small samples ($n\le 5$), where no analytical expression of the optimal seed is known. 
For mixed states the expression of $\Gamma_\uparrow$ and $\Omega_m$ can be found in Sec.~Methods, Eqs.~(\ref{jA - jC 2}) through~(\ref{SDP methods 2}).

\begin{figure}[ht]
\includegraphics[scale=0.75]{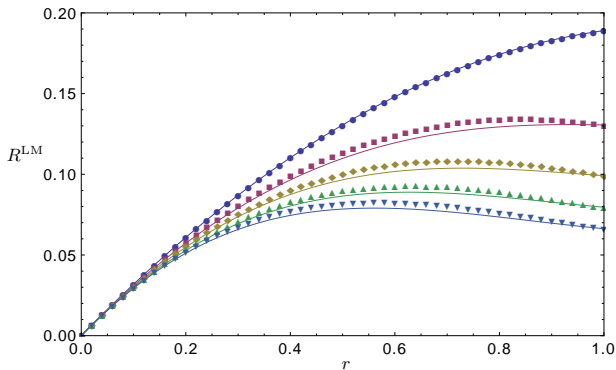}
\caption[]{\label{fig1}(Color online) Excess risk $R^{\rm LM}$ (points) and its corresponding lower bound $R^{\rm opt}$ (lines), both as a function of the purity $r$, and for values of $n$ ranging from 1 to 5 (from top to bottom).}
\end{figure}

%
%
%
Our results are shown in Fig.~\ref{fig1}, where we plot $R^{\rm LM}$ (shaped dots) and the lower bounds given by $R^{\rm opt}$ (solid lines) as a function of the purity $r$ for up to~$n=5$. We note that the excess risk of the optimal LM is always remarkably close to the absolute minimum provided by the optimal programmable machine and in the worst case ($n=2$) it is only $0.4\%$ larger.
For $n=1$ we see that $R^{\rm LM}=R^{\rm opt}$ for any value of~$r$. This must be the case since for a single qubit in $A$ and $C$ one has $j_A=j_C=1/2$, and Eq.~(\ref{ja=jc}) holds.
%
%
%


We now turn to robustness against statistical fluctuations in the number of states of each type produced by the source. In a real scenario one has to expect that~$j_A=n_A/2\not=n_C/2=j_C$, $n_A+n_B=2n$. Hence,~$\Gamma_\uparrow$ has the general form~(\ref{jA - jC 2}), which gives us a hint that our choice $\Omega=\Omega_{m=0}$ may not be optimal for finite~$n$. This has been confirmed by numerical analysis using the same SDP approach discussed above. Here, we show that the asymptotic performance (for large training samples) of the optimal LM, however, is still the same as that of the optimal programmable discriminator running under the same conditions (mixed states and statistical fluctuations in $n_{A/C}$).  

Asymptotically, a real source for the problem at hand will typically produce~$n_{A/C}=n\pm \delta\sqrt{n}$ mixed copies of each type. 
In Sec.~Methods, it is shown that the relation~(\ref{s-s asymp}) still holds in this case if $n$ is large. It reads
\begin{equation}\label{ja=jc asymp}
\sigma^n_{0,\xi}-\sigma^n_{1,\xi}  \simeq r \left(1-{1-r\over n r^2}\right)  \left(\sigma^{rn}_{0}-\sigma^{rn}_{1}\right)
\end{equation}
($\delta$ first appears at order $n^{-3/2}$). Hence, the effect of both statistical fluctuations in $n_{A/C}$ and noise (already considered above) is independent of the machine used for quantum classification (i.e., it is the same for LM, programmable machines, E\&D, \dots). In particular, the relation~(\ref{rlm=ropt}), $R^{\rm LM}=R^{\rm opt}$, between the excess rate of the optimal LM and its absolute limit given by the optimal programmable discriminator still holds asymptotically, which proves robustness.

%
%
%
%
%

To illustrate this, let us consider the effect of statistical fluctuations in~$n_{A/C}$ for pure states.
The optimal programmable machine for arbitrary $n_A$, $n_B$ and $n_C$ was presented in~\cite{Sentis}. The error probability for the case at hand  ($n_B=1$) is in Sec.~Methods.  From its asymptotic expansion when $n_A$ and $n_C$ are both large one readily has
\begin{equation}\label{opt asym}
R^{\rm opt} = \frac{1}{6} \left(\frac{1}{n_A}+\frac{1}{n_C}\right)+\dots\,.
\end{equation}
We see that when~$n_{A/C}=n\pm \delta\sqrt{n}$ (i.e., when statistical fluctuations in $n_{A/C}$ are taken into account) one still has~$R^{\rm opt}\simeq1/(3n)\simeq R^{\rm LM}$. 

\section*{Discussion}

We have presented  a \emph{supervised} quantum learning machine that classifies a single qubit prepared in a pure but otherwise unknown state after it has been trained with a number of already classified qubits.
Its performance attains the absolute bound given by the optimal programmable discrimination machine. 
This learning machine does not require quantum memory and can also be reused without retraining, 
which may save a lot of resources.
The machine has been shown to be robust against noise and statistical fluctuations in the number of states of each type produced by the source.
For small sized training sets the machine is very close to optimal, attaining an excess risk that is larger than the absolute lower limit by at most $0.4\%$. In the absence of noise and statistical fluctuations, the machine attains optimality for {\em any} size of the training set.


One may rise the question of whether or not the separated measurements on the training set and data qubit can be reversed in time; in a classical scenario where, e.g., one has  to identify one of two faces based on a stack of training portraits, it is obvious that, without memory limitations, the order of training and data observation can be reversed (in both cases the final decision is taken based on the very same information).
We will briefly show that this is not so in the quantum world.
In the reversed setting, the machine first performs a measurement~${\mathscr D}$, with each element of rank one~$u_\mu[\,\uparrow\,] u^\dagger_\mu$,  and stores the information (which of the possible outcomes is obtained) in the classical memory to control the measurement to be performed on the training set in a later time. 
The probability of error conditioned to one of the outcomes, say $\uparrow$, is given by the Helstrom formula $P_e^{\uparrow}=(1-\trnorm{\!\!\Gamma_\uparrow\!\!}\!\!/2)/2$, where $\Gamma_\uparrow$ is defined in Eq.~(\ref{def GammaUp}). Using Eq.~(\ref{JA-JC}) one has $\trnorm{\!\Gamma_\uparrow\!}=d^{-2}_n d^{-1}_{n+1}\sum_{m,m'}|m-m'|=n/[3(n+1)]$. The averaged error probability is then
\begin{equation}\label{PeLMback}
P^{\stackrel{{\rm LM}}{\mbox{\tiny$\leftarrow$}}}_e=\frac{1}{2} \left(1-\frac{1}{6}\frac{n}{n+1}\right).
\end{equation}
In the limit of infinite copies we obtain $P^{\stackrel{{\rm LM}}{\mbox{\tiny$\leftarrow$}}}_e\simeq 5/12$, which is way larger than $P^{\rm LM}_e\simeq 1/6$. The same minimum error probability of~Eq.~(\ref{PeLMback}) can be attained by performing a Stern-Gerlach measurement on the data qubit, which requires just one bit of classical memory. This is all the classical information that we can hope to retrieve from the data qubit, in agreement with Holevo's bound~\cite{Holevo bound}. This clearly limits the possibilities of a correct classification ---very much in the same way  as in face identification with limited memory size.
In contrast, the amount of classical information ``sent forward" in the optimal learning machine goes as the logarithm of the size of the training sample.  This asymmetry also shows that despite the separability of the measurements, non-classical correlations between the training set and the data qubit play an important role in quantum learning. 


Some relevant generalizations of this work to, e.g., higher dimensional systems and arbitrarily unbalanced training sets, remain an open problem.  Another challenging problem with direct practical applications in quantum control and information processing is the extension of this work to \emph{unsupervised} machines, where no human expert classifies the training sample.


\section*{Methods}

\subsection*{\boldmath Block-diagonal form of $\rho^{\otimes n}$}

The state $\rho^{\otimes n}$ of $n$ identical copies of a general qubit state $\rho$ with purity~$r$ and Bloch vector $r\vec{s}$, has a block diagonal form in the basis of the total angular momentum given by
\begin{equation*}
\rho^{\otimes n} = \sum_{j} p^n_j  \rho_{j} \otimes {\idd_j\over \nu^n_j}.
\end{equation*}
Here $j=0\,(1/2),\hdots,n/2$ if $n$ is even (odd), $\idd_j$ is the identity in the multiplicity space ${\mathbb C}^{\nu^n_j}$, of dimension $\nu^n_j$ (the multiplicity of the representation with total angular momentum $j$), where
\begin{equation*}
\nu^n_j= 
\begin{pmatrix}
n \\
n/2-j
\end{pmatrix}
\frac{2j+1}{n/2+j+1} .
\end{equation*}
The normalized state $\rho_{j}$, which is supported on the representation  subspace $\mathscr{S}_{j}={\rm span}\{\ket{j,m}\}$ of dimension~$2j+1=d_{2j}$,~is
\begin{equation*}
\rho_{j} = U_{\scrvec  s}\left( \sum_{m=-j}^{j} a^j_m\; [j,m]\right)U_{\scrvec  s}^{\dagger}  ,
\end{equation*}
where
\begin{equation}
a^j_m=\frac{1}{c_j} \left(\frac{1-r}{2}\right)^{j-m} \left(\frac{1+r}{2}\right)^{j+m} ,
\label{a_m}
\end{equation}
and
\begin{equation*}
c_j = \frac{1}{r} \left\{\left(\frac{1+r}{2}\right)^{2j+1}\!\!\!-\left(\frac{1-r}{2}\right)^{2j+1}\right\} ,
\end{equation*}
so that $\sum_{m=-j}^j a^j_m=1$,
and we stick to our shorthand notation $[\,\cdot\,]\equiv|\,\cdot\,\rangle\langle\,\cdot\,|$, i.e., $[j,m]\equiv \ketbrad{j,m}$.
The measurement on $\rho^{\otimes n}$ defined by the set of projectors on the various subspaces~$\mathscr{S}_{j}$ will produce $\rho_{j}$ as a posterior state with probability
\begin{equation*}
p^n_j = \nu^n_j c_j \left(\frac{1-r^2}{4}\right)^{n/2-j} .
\end{equation*}
One can easily check that $\sum_j p^n_j=1$. 

In the large $n$ limit, we can replace $p^n_j$ for a continuous probability distribution $p_n(x)$ in $[0,1]$, where $x=2j/n$. Applying Stirling's approximation to $p_j$ one obtains:
\begin{equation*}
p_n(x)\simeq \sqrt{\frac{n}{2 \pi}}\frac{1}{\sqrt{1-x^2}} {x(1+r)\over r(1+x)}\;\mathrm{e}^{- n H(\frac{1+x}{2}\parallel\frac{1+r}{2})},
\end{equation*}
where $H(s\parallel t)$ is the (binary) relative entropy
\begin{equation*}
H(s\parallel t)=s \log\frac{s}{t}+ (1-s)\log\frac{1-s}{1-t} .
\end{equation*}
The approximation is valid for $x$ and $r$ both  in the open unit interval~$(0,1)$. For non-vanishing $r$, $p_n(x)$ becomes a Dirac delta function peaked at $x=r$, $p_\infty(x)=\delta(x-r)$, which corresponds to~$j=nr/2$.
\\

\subsection*{\boldmath Covariance and structure of $\mathscr{L}$}

We start with a POVM element of the form $\bar{E}_0=\int du\,U\, E_0\, U^{\dagger}$. Since $D_\mu$ must be a rank-one projector, it can always be written as $D_\mu = u_\mu\, [\,\uparrow\,] \,u_\mu^\dagger$ for a suitable SU(2) rotation $u_{\mu}$. Thus,
\begin{equation*}
{\bar E}_0 = \sum_\mu\int du \left(U_{AC} L_\mu U_{AC}^\dagger\right)\otimes   \left(u u_\mu[\,\uparrow\,]u_\mu^\dagger u^\dagger \right).
\end{equation*}
We next use the invariance of the Haar measure $du$ to make the change of variable: $u\,u_{\mu}\rightarrow u'$ and, accordingly,  $U_{AC} \rightarrow U'_{AC} U^{\dagger}_{\mu \, {AC}}$. After regrouping terms we have
\begin{eqnarray*}
{\bar E}_0\!&=&\! \sum_\mu\! \int\!\! du' \left( U'_{AC} U^\dagger_{\mu\, {AC}}  L_\mu U_{\mu\, {AC}}  U'{}_{AC}^\dagger\right)\! \otimes \! \left( u'[\,\uparrow\,]u'{}^\dagger\right)  \nonumber \\
&=&\!\! \int\!\! du' \!\left[U'_{AC}\!\left(\! \sum_\mu U^\dagger_{\mu\, {AC}}  L_\mu U_{\mu\, {AC}}\!\!\right)  U'{}_{AC}^\dagger\right]\!\otimes\!  \left(u'[\,\uparrow\,]u'{}^\dagger \right)\nonumber \\
&=& \int\!\! du \left(U_{AC}\, \Omega\, U^\dagger_{AC}\right)\otimes \left(u[\,\uparrow\,] u^{\dagger}\right) ,
\end{eqnarray*}
where we have defined \mbox{${\Omega} =\sum_\mu U^\dagger_{\mu\,AC}L_\mu U_{\mu\,AC}\ge 0$.}
The POVM element $\bar{E}_1$ is obtained by replacing~\mbox{$[\,\uparrow\,]$} by~\mbox{$[\,\downarrow\,]$} in the expressions above. From the POVM~condition $\sum_\mu L_\mu=\id_{AC}$ it immediately follows that~$\int du \, U_{AC} \,{\Omega} \,U^\dagger\kern-.3em{}_{AC}= \id_{AC}$, where $\id_{AC}$ is the identity on the Hilbert space of the TS, i.e., $ \id_{AC}=\id_A \otimes \id_C$. Therefore ${\mathscr L}=\{U_{AC} \,{\Omega} \,U^\dagger\kern-.3em{}_{AC}\}_{\rm SU(2)}$ is a covariant POVM. The positive operator $\Omega$ is called the seed of the covariant POVM~$\mathscr L$. 

Now, let $u_z(\varphi)$ be a rotation about the $z$-axis, which leaves $[\,\uparrow\,]$ invariant. By performing the change of variables $u\rightarrow u'u_z(\varphi)$ [and $U_{\,AC}\rightarrow U'_{AC}U_{\,zAC}(\varphi)$] in the last equation above, we readily see that $\Omega$ and $U_{\,zAC}(\varphi)\,\Omega\,U^\dagger_{\,zAC}(\varphi)$ both give the same average operator $\bar E_0$ for any $\varphi\in[0,4\pi)$. So, its average over $\varphi$,
\begin{equation*}
\int_0^{4\pi} {d\varphi\over4\pi} U_z(\varphi)\,\Omega\, U_z^\dagger(\varphi) ,
\end{equation*}
can be used as a seed w.l.o.g., where we have dropped the subscript $AC$ to simplify the notation. Such a seed is by construction invariant under the group of rotations about the $z$-axis (just like $[\,\uparrow\,]$) and, by Schur's lemma, a direct sum of operators with well defined magnetic number. Therefore, in the total angular momentum basis for~$AC$, we can always choose the seed of $\mathscr L$ as
%
\begin{equation*}
 \Omega=\sum_{m=-n}^n\Omega_m; 
\quad \Omega_m\ge0 .
\end{equation*}
%
%
The constrain~(\ref{omega_m}) follows from the POVM condition ${\id}_{AC}=\int du\, U\,\Omega\, U^\dagger$ and Schur's lemma.
The result also holds if~$A$ and $C$ have different number of copies (provided they add up to $2n$). 
It also holds for mixed states.
\\

\subsection*{Wigner's 6j-symbols}

Let us consider three angular momenta $j_1, j_2, j_3$ that couple to give a total $J$. Note that there is no unique way~to carry out this coupling; we might first couple $j_1$ and~$j_2$ to give a resultant $j_{12}$, and couple this to $j_3$ to give~$J$, or alternatively, we may couple $j_1$ to the resultant~$j_{23}$ of coupling $j_2$ and $j_3$. Moreover, the intermediate couplings can give in principle different values of~$j_{12}$ or~$j_{23}$ which, when coupled to $j_3$ or $j_1$, end up giving the same value of $J$. All these possibilities lead to linearly independent states with the same $J$ and $M$, thus they must be distinguished by specifying the intermediate angular momentum and the order of coupling. There exists a unitary transformation that maps the states obtained from the two possible orderings of the coupling; Wigner's 6j-symbols~\cite{edmonds}, denoted in the next equation by $\{ \, {}^{\cdots}_{\cdots} \, \}$, provide the coefficients of this transformation:
\begin{eqnarray*}
&&\langle(j_1\,j_2)j_{12},j_3;J,M|j_1,(j_2\,j_3)j_{23};J,M\rangle \\[.5em]
&&=\! (-1)^{j_1+j_2+j_3+J}\! \sqrt{\!(2j_{12}+1)(2j_{23}+1)}
\!\left\{
\begin{array}{ccc}
\!j_1\! &\! j_2\! &\! j_{12}\!\! \\
\!j_3\! &\! J\! &\! j_{23}\!\!
\end{array}
\right\}  .
\end{eqnarray*}
Note that this overlap is independent of $M$. For the proof of optimality of the LM, we couple subsystems~$A$,~$B$ and~$C$ in two ways: $A(CB)$ and $(AC)B$ to produce the states $\ket{j_A,(j_C\,j_B)j_{CB};J,M}$ and $\ket{(j_A\,j_C)j_{AC},j_B;J,M}$, which we denote by $|J,M\rangle_{\Xyz ACB}$ and $|J,M\rangle_{\xyZ ACB}$ respectively for short. The various angular momenta involved are fixed to \mbox{$j_A=j_C=\mbox{\footnotesize${n\over2}$}$}, $j_B=\half$, $j_{AC}=j$, $j_{CB}=\mbox{\footnotesize${n\over2}$}+\half$, whereas $J=j\pm\half$. With these values, the overlaps we need are given by
\begin{equation*}
  {{}_{\Xyz ACB}\kern-.1em\langle j\pm\half,\half|j\pm\half,\half\rangle_{\xyZ ACB}}=
  \sqrt{\frac{
  {n+\mbox{\footnotesize${3\over2}$}\pm(j+\half)}}{2(n+1)}}.
\end{equation*}
%

\subsection*{Derivation of Eqs.~(\ref{ja=jc}) and~(\ref{ja=jc asymp})}

Let us start with the general case where $\xi=\{j,j'\}$. To obtain $\sigma^n_{0,\xi}$ we first write Eqs.~(\ref{sigmaxi1}) and~(\ref{sigmaxi2}) as the SU(2) group integrals 
%
\begin{eqnarray*}
\sigma^n_{0,\xi}\!&=&\!\!\int\!\!du \, U_{AB} \!\left( \sum_{m=-j}^j\!\!\! a^j_m [j,m]_A\! \otimes\!\rho^B_0\! \right) \! U_{AB}^{\dagger} \\
\!\!& \otimes&\!\! \int\!\! du' \, U'_C\! \left(\sum_{m=-j'}^{j'}\!\! a^{j'}_m [j',m]_C\! \right) U'^{\dagger}_C \, ,
\end{eqnarray*}
where $a^j_m$ is given in Eq.~(\ref{a_m}),
%
%
$\rho^B_0$ is the mixed state~$\rho_0$, Eq.~(\ref{rho mixed}), of the qubit~$B$. We next couple $A$ with $B$ (more precisely, their subspaces of angular momentum~$j$) using the Clebsch-Gordan coefficients 
\begin{eqnarray*}
|\!\braket{j+\half,m+\half}{j,m;\half,\half}\!|^2 &=&\frac{j+ m+1}{2j+1} \, , \\
|\!\braket{j-\half,m+\half}{j,m;\half,\half}\!|^2 &=&\frac{j- m}{2j+1} \, .
\end{eqnarray*}
The resulting expressions can be easily integrated using Schur lemma. Note that the integrals of crossed terms of the form $\ketbra{j,m}{j',m}$ will vanish for all $j\neq j'$. We~readily obtain
\begin{equation*}
\sigma^n_{0,\xi}\!=\!\! \!\sum_{m=-j}^j\!\! \!a^j_m\!\!  \left(\! \frac{j\!+\!1\!+\!m r}{d_{2j}}\, {\id_{2j+1}^{AB} \over  d_{2j+1}}+ \frac{j\!-\!m r}{d_{2j}}\,{ \id_{2j-1}^{AB}\over  d_{2j-1} } \!\right)\! \otimes\! \frac{ \id_{2j'}^C}{d_{2j'}} ,
\end{equation*}
%
%
%
where 
$\id_{2j}$ is the projector on~$\mathscr{S}_{j}$ and $d_{2j}=2j+1=\dim\mathscr{S}_{j}$. The superscripts attached to the various projectors specify the subsystems to which they refer. These projectors are formally equal to those used in Eq.~(\ref{sigma states}) (i.e., $\id_{2j}$ projects on the fully symmetric subspace of $2j$ qubits) and, hence, we stick to the same notation.   Note that $\tr\sigma^n_{0,\xi}=1$, as it should be. 

We can further simplify this expression by introducing $\langle\hat{J}_z\rangle_j=\sum_m m\, a^j_m$, i.e.,  the expectation value of the $z$-component of the total angular momentum in the state~$\rho_{j}$ (i.e., of $\id_{2j}\hat{J}_z\id_{2j}$ in the state~$\rho^{\otimes n}_{0/1}$)  for a Bloch vector~$r {\vec{z}} $:
\begin{equation*}
\sigma^n_{0,\xi}\!=\!\!  \left(\! \frac{j\!+\!1\!+\! r\langle\hat{J}_z\rangle_{\!j}}{d_{2j}}\, {\id_{2j+1}^{AB} \over  d_{2j+1}}+ \frac{j\!-\!r\langle\hat{J}_z\rangle_{\!j}}{d_{2j}}\,{ \id_{2j-1}^{AB}\over  d_{2j-1} } \!\right)\! \otimes\! \frac{ \id_{2j'}^C}{d_{2j'}} .
\end{equation*}
%
%
Using the relation
\begin{equation*}
\id_{2j-1}^{AB}=\id_{2j}^{A}\otimes \id^B_{1}-\id_{2j+1}^{AB},
\end{equation*}
%
and
$(j+1)/d_{2j+1}=j/d_{2j-1}=1/2$, we can write
\begin{equation}
\sigma^n_{\!0,\xi}\!=\!\!  \left(\! \!{r\langle\hat{J}_{\!z}\rangle_{\!j}\over j}{\id_{2j+1}^{AB} \over  d_{2j+1}}
\!+\! \frac{j\!-\!r\langle\hat{J}_{\!z}\rangle_{\!j}}{j}\,{\id_{2j}^{A}\over d_{2j}}\!\otimes\! {\id^B_{1}\over2}\!\!\right)\! \otimes\! \frac{ \id_{2j'}^C}{d_{2j'}}.\label{new sigma0xi}
\end{equation}
Similarly, we can show that
\begin{equation}
\sigma^n_{\!1,\xi}\!=\! \frac{ \id_{2j}^A}{d_{2j}}\! \otimes\!\! \left(\!\! \frac{r\langle\hat{J}_{\!z}\rangle_{\!j'}}{j'}{\id_{2j'\!+1}^{BC} \over  d_{2j'\!+1}}\!+ \!\frac{j'\!\!-\!r\langle\hat{J}_{\!z}\rangle_{\!j'}}{j'}\, \!{\id^B_{1}\over2}\!\!\otimes\! {\id^{C}_{2j'}\over d_{2j'}}\! \!\right)\! \!.
\label{new sigma1xi}
\end{equation}
Therefore, if $j'=j$,
\begin{equation*}
\sigma^n_{0,\xi}\!-\!\sigma^n_{1,\xi}\!=\!{r\langle\hat{J}_z\rangle_{\!j}\over j}\!\!\left({\id^{AB}_{2j+1}\over d_{2j+1}}\!\otimes\!{\id^{C}_{2j}\over d_{2j}}\!-\!{\id^{A}_{2j}\over d_{2j}}
\!\otimes\! {\id^{BC}_{2j+1}\over d_{2j+1}}\right)\!.
\end{equation*}
Comparing with Eq.~(\ref{sigma states}), 
the two terms in the second line can be understood as 
the average states for a number of $2j$ pure qubits, i.e., as $\sigma^{2j}_0$ and  $\sigma^{2j}_1$~respectively. Hence, if $\xi=\{j,j\}$ we have the relation
\begin{equation*}
\sigma^n_{0,\xi}-\sigma^n_{1,\xi}  = \frac{r \mean{\hat{J}_z}_j}{j}   \left(\sigma^{2j}_{0}-\sigma^{2j}_{1}\right) ,
\end{equation*}
which is Eq.~(\ref{ja=jc}). 
It is important to emphasize that this equation is exact (i.e.,~it holds for any value of $j$, $n$ and~$r$) and bears no relation whatsoever to measurements (i.e., it is an algebraic identity between the various operators involved).

In the asymptotic limit, for $n_A$ and $n_C$ of the form $n_{A/C}\simeq n\pm b n^{a}$,  $n\gg 1$, $a<1$, the probabilities $p^n_j$ and $p^n_{j'}$ are peaked at $j\simeq r n_A/2$ and~$j'\simeq r n_C/2$, as was explained above. Hence, only the average state components $\sigma^n_{0/1,\xi}$ with $\xi=\{j,j'\} $ such that $j \simeq (r/2) n(1+b n^{a-1})$ and $j' \simeq (r/2) n(1-b n^{a-1})$ are important. From~Eqs.~(\ref{new sigma0xi}) and~(\ref{new sigma1xi}) it is straightforward to obtain

\begin{equation*}
\sigma^n_{0,\xi}-\sigma^n_{1,\xi}  \simeq r \left(1-{1-r\over n r^2}\right)  \left(\sigma^{rn}_{0}-\sigma^{rn}_{1}\right) +o(n^{-1}),
\end{equation*}
where we have used that~\cite{BEJRE} $\langle\hat{J}_z\rangle_j\simeq j-(1-r)/(2r)$ up to exponentially vanishing terms. This relation, for the particular value of $a=1/2$, is used in the proof of robustness, Eq.~(\ref{ja=jc asymp}).
\\

\subsection*{\boldmath Calculation of $\Gamma_\uparrow$}

Here we calculate $\Gamma_{\uparrow,\xi}\!=\!\tr_{\!B}\{[\,\uparrow\,] (\sigma^n_{0,\xi}-\sigma^n_{1,\xi})\}$, where the average states are defined in Eqs.~(\ref{sigmaxi1}) and~(\ref{sigmaxi2}), and explicitly given in~Eqs.~(\ref{new sigma0xi}) and~(\ref{new sigma1xi})  for~$\xi=\{j,j'\}$. Let us first calculate the conditional state $\tr_{\!B}([\,\uparrow\,] \sigma^n_{0,\xi})$. For that, we need to express $\id^{AB}_{2j+1}=\sum_m [j+\half,m]$ in the original product basis $\{\ket{j_A,m_A}\otimes\ket{\uparrow/\downarrow}\}$.
Recalling the Clebsch-Gordan coefficients $|\braket{\half,\half; j, m}{j+\half,m+\half}|^2=(j+m+1)/(2j+1)$, one readily obtains
\begin{equation*}
\tr_{\!B}\left([\,\uparrow\,] {\id^{AB}_{2j+1}\over d_{2j+1}}\right)=\!\!\sum_{m=-j}^{j}\!\frac{j\!+\!1\!+\!m}{2(j+1)d_{2j}}[j,m]_A ,
\end{equation*}
which can be written as
\begin{equation*}
\tr_{\!B}\left([\,\uparrow\,] {\id^{AB}_{2j+1}\over d_{2j+1}}\right)=
{1\over2}\left({\id^A_{2j}\over d_{2j}}+{1\over d_{2j}}{\hat{J}^A_z\over j+1}\right) ,
\end{equation*}
where $\hat J^{A}_z$ is the $z$ component of the total angular momentum operator acting on subsystem $A$.
An analogous expression is obtained for  $\tr_{\!B}\left([\,\uparrow\,] \id^{BC}_{2j'+1}\right)$. Substituting in Eqs.~(\ref{new sigma0xi}) and~(\ref{new sigma1xi}) and subtracting the resulting expressions, one has $\Gamma_\uparrow=\sum_{\xi}p^n_\xi \Gamma_{\!\uparrow,\xi}$, with
\begin{equation}\label{jA - jC 2}
\Gamma_{\!\uparrow,\xi}\!=\!{1\over2d_{2j_A}d_{2j_C}}\!\!
\left(\!\!
{r\langle\hat{J}_z\rangle_{j_A}\over j_A}{\hat{J}^A_z\over j_A+1}
\!-\!
{r\langle\hat{J}_z\rangle_{j_C}\over j_C}{\hat{J}^C_z\over j_C+1}
\!\!
\right)\!\!,
\end{equation}
where we have written $\xi=\{j_A,j_C\}$, instead of $\xi=\{j,j'\}$ used in the derivation. For pure states,
$r=1$, 
$j_A=j_C=n/2$, $\langle\hat{J}_z\rangle_{n/2}=n/2$, and we recover Eq.~(\ref{JA-JC}).

In order to minimize the excess risk using SDP, we find it convenient to write Eq.~(\ref{Delta SDP})  in the form
\begin{equation}\label{SDP methods 1}
\Delta^{\rm LM}=2\max_{\{\Omega_{m,\xi}\}} \sum_\xi p^n_\xi \tr(\Gamma_{\uparrow,\xi}\Omega_{m,\xi}),
\end{equation}
where we recall that $m=m_{AC}=m_A+m_C$, and  we assumed w.l.o.g. that the seed of the optimal POVM has the block form~$\Omega_{m}=\sum_\xi\Omega_{m,\xi}$. The POVM condition, Eq.~(\ref{omega_m}) must now hold on each block, thus for~\mbox{$\xi=\{j_A,j_C\}$}, we must impose that
\begin{equation}\label{SDP methods 2}
\sum_{m=-j}^j\!\!\langle j,m|\Omega_{m,\xi}|j,m\rangle\!=\!2j\!+\!1,\,|j_A\!-\!j_C|\!\le\! j\!\le\! j_A\!+\!j_C.
\end{equation}


\subsection*{Programmable machine for unbalanced training sets}

The minimum error probability of the optimal programmable machine with a number of $n_A$, $1$ and $n_C$ copies ($n_A>n_C$) in ports $A$, $B$ and $C$ respectively, is~\cite{Sentis}
\begin{eqnarray*}
P_e^{\rm opt} \! &=&\! {1\over4}\! \left\{ \!1\!+\!{D_0\over D_1}\!-\!{D_0\!+\!D_1\over D_0D_1}\sum_{k=0}^{n_C}
(n_A\!-\!n_C\!+\!2k\!+\!2) \right. \\[.5em]
\!&\times&\! \left.
\sqrt{
1\!-\!4{D_0D_1\over(D_0\!+\!D_1)^2}\frac{(n_A\!-\!n_C\!+\!k\!+\!1) (k\!+\!1)}{(n_A+1)(n_C+1)}
}
\right\} ,
\end{eqnarray*}
where $D_0 = (n_A+2)(n_C+1)$ and $D_1= (n_A+1)(n_C+2)$ are the dimensions of the average states $\sigma_{0/1}$. The asymptotic form of this expression when $n_A$ and $n_C$ are both  very large can be easily  derived using Euler-Maclaurin's summation formula.  The result up to subleading order is
\begin{equation*}
P_e^{\rm opt} \simeq \frac{1}{6} \left(1+\frac{1}{n_A}+\frac{1}{n_C}\right),
\end{equation*}
which leads to Eq.~(\ref{opt asym}).
%

\bigskip

\section*{Acknowledgements}

We acknowledge financial support from: ERDF: European Regional Development Fund;
the Spanish MICINN, through contract FIS2008-01236, FPI Grant No. BES-2009-028117 (GS)
and (EB) PR2010-0367; and from the Generalitat de
Catalunya CIRIT, contract  2009SGR-0985.

%
%

\end{document}